\documentclass[12pt]{iopart}
\usepackage{graphicx}
\usepackage{hyperref} 
\usepackage{xcolor}
\usepackage[numbers,sort&compress]{natbib}
\usepackage{caption}
\usepackage{subcaption}

\newcommand{\ket}[1]{$|#1\rangle$}
\newcommand{\yso}{Y$_2$SiO$_5$}
\newcommand{\euyso}{$^{151}$Eu$^{3+}$:Y$_2$SiO$_5$}
\newcommand{\pryso}{Pr$^{3+}$:Y$_2$SiO$_5$}
\newcommand{\eu}{$^{151}$Eu$^{3+}$}

\begin{document}

\title{Efficient and reversible optical-to-spin conversion for solid-state quantum memories}

\author{Jingjing Chen and Mikael Afzelius}

\address{Department of Applied Physics, University of Geneva, 1205 Geneva, Switzerland}

\ead{mikael.afzelius@unige.ch}
\vspace{10pt}
%\begin{indented}
%\item[]July 2024
%\end{indented}

\begin{abstract}
Long-duration and efficient quantum memories for photons are key components of quantum repeater and network applications. To achieve long duration storage in atomic systems, a short-lived optical coherence can be mapped into a long-lived spin coherence, which is the basis for many quantum memory schemes. In this work, we present modeling and measurements of the back-and-forth, i.e. reversible, optical-to-spin conversion for an atomic frequency comb memory. The AFC memory is implemented in \euyso{} with an applied magnetic field of 231 mT, which allows lifting Zeeman transition degeneracy which otherwise cause time-domain interference in the optical-to-spin conversion. By optimizing the conversion using the developed simulation tool, we achieve a total efficiency of up to 96\%, including the spin echo sequence and spin dephasing, for a storage time of 500 $\mu$s. Our methods and results pave the way for long-duration storage of single photon states in \euyso{} with high signal-to-noise, at the millisecond timescale.
\end{abstract}

%\maketitle

\section{Introduction}
\label{sec:intro}

Quantum memories for optical photons have applications in various quantum information schemes \cite{Bussieres2013,Heshami2016}, in particular for quantum communication via quantum repeaters \cite{Duan2001,Sangouard2011}. In quantum repeater schemes based on independent photonic sources of entangled states and quantum memories, long duration storage is typically based on mapping the input state from the optical domain into the spin domain of an atomic system, which can be achieved by various methods specific to the physical implementation of the quantum memory \cite{Cirac1997,Lukin2003,Nunn2007,Afzelius2009a,Hosseini2012}.

In this article, we present a study of the modeling, optimization and measurement of the mapping of light from the optical to the spin domain in a solid-state quantum memory, based on the atomic frequency comb (AFC) scheme \cite{Afzelius2009a}. An AFC spin-wave memory \cite{Afzelius2010, Jobez2015, Gundogan2015, Businger2020, Ortu2021, Rakonjac2021, Ma2021} can achieve multimode storage \cite{Nunn2008,Ortu2022}, a key requirement for long-distance quantum repeaters, in combination with long-duration storage by an optical-to-spin mapping. In practice, this mapping requires the application of an efficient optical population inversion pulse over the entire AFC bandwidth \cite{Minar2010}. Futhermore, the application of population inversion pulses on the spin transition in the radio frequency \cite{Jobez2015,Ortu2021} or microwave \cite{Businger2020} domain can extend the spin storage time by several orders of magnitude, through dynamical decoupling of the spin coherence \cite{Holzaepfel2020, Ortu2021, Ma2021}, provided that the pulses are efficient over the entire spin broadening \cite{Holzaepfel2020}. It has been a challenge, however, to design optical and spin population inversion pulse sequences for achieving a reversible optical-to-spin mapping with an efficiency approaching unity, which is one of the goals of this article. 

We here focus on an AFC spin-wave memory in \euyso{} \cite{Jobez2015,Holzaepfel2020,Ortu2021,Ma2021}. The Eu$^{3+}$ ion is a non-Kramers ion having only nuclear hyperfine levels in the relevant electronic states \cite{Koenz2003}, and it is particularly challenging in this context due to its low optical and spin transition dipole moments \cite{ZambriniCruzeiro2018a,Ma2018}. Yet, Eu-doped \yso{} is highly interesting thanks to its uniquely long spin coherence times, which can reach hours in some experimental configurations under magnetic field \cite{Zhong2015,Ma2021}. To overcome the weak transition moments, we use optical and spin inversion pulses with smooth amplitude envelopes that are frequency chirped over the relevant AFC and spin bandwidths, which become adiabatic in the asymptotic limit of long pulses. Recently, we also identified that weakly non-degenerate Zeeman doublets in \euyso{} , due to the unavoidable presence of weak magnetic bias fields, cause complex modulations of the memory efficiency \cite{Etesse2021}, due to the excitation of near-degenerate transitions by the optical and spin pulses. To counter this, we apply a moderate magnetic field of about 230 mT in order to completely split the optical/spin transitions, such that each addressed state in the protocol is a single state. 

To gain insight into the limitations of our optical and spin control sequences, and to optimize their efficiencies, we develop numerical 2-dimensional (2-d) Bloch-equation models accounting for most experimental parameters, including the Gaussian intensity distribution of the optical modes and their overlap in a crossed-beam configuration, the AFC bandwidth and spin linewidth. Our main results show that by carefully optimizing the sequences, we reach a total back-and-forth optical-to-spin efficiency of up to $(96 \pm 1)$ \%, including all optical and spin inversion pulses and spin dephasing, in the limit of long adiabatic pulses.

The methods presented in this paper are also highly relevant to AFC spin-wave quantum memories in praseodymium-doped crystals \cite{Gundogan2015, Rakonjac2021}, which is an analogous non-Kramers system with the same hyperfine structure, although with somewhat larger transition dipole moments in general. More generally, many of our results are applicable to any quantum memory relying on optical and spin inversion pulses that need to efficiently address bandwidths that are challenging to excite with simple $\pi$-pulses, such as the the noiseless photon-echo (NLPE) \cite{Ma2021a} or the rephased amplified spontaneous emission (RASE) \cite{Ledingham2010} protocols.

\section{The AFC spin-wave memory protocol}
\label{sec:AFC-prot}

\begin{figure}[h]
    \includegraphics[width=0.8\linewidth]{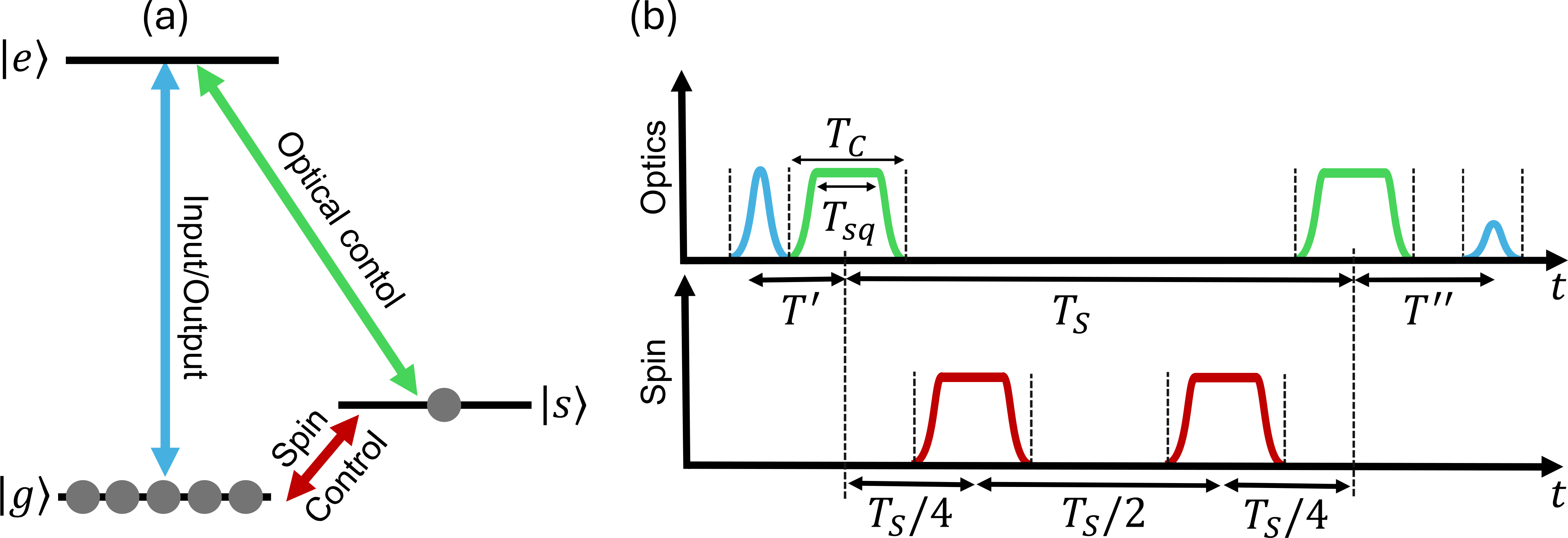}
    \centering
    \caption{(a) Schematic energy level diagram and transitions involved in the AFC spin-wave quantum memory. Shown are the transitions for the input/output modes (blue), and the optical (green) and spin (red) population inversion pulses denoted ''optical control'' and ''spin control'', respectively. (b) Time sequence of the entire memory scheme. The first optical control pulse is applied a time $T'$ after the input pulse and the time separation between the optical control pulses is $T_S$. The output mode will appear a time $T''$ after the second optical control pulse, such that $T'+T'' = 1/\Delta$, where $\Delta$ is the AFC periodicity.}
    \label{fig:AFC_scheme}
\end{figure}

The AFC spin-wave memory has been described in detail elsewhere \cite{Afzelius2009a,Afzelius2010,Ortu2022}, here we will briefly recall the main features of the protocol and define the reversible optical-to-spin efficiency, the main quantity of interest in this article. The core feature of all AFC memories is a periodic structure of sharp and highly absorbing peaks (teeth), which can be created through optimized spectral hole burning (SHB) techniques \cite{Jobez2016,Businger2022}. If the periodicity of the comb is $\Delta$ in frequency space, then the absorption of an input mode, with a spectrum fully absorbed by the AFC bandwidth, will cause an output echo emission after a delay $1/\Delta$ \cite{Mitsunaga1991,Merkel1996,Afzelius2009a}. The efficiency of this AFC echo process is given by the optical depth on the input/output transition \ket{g}--\ket{e}, see Fig. \ref{fig:AFC_scheme}a, and the shape of the comb structure \cite{Afzelius2009a,Bonarota2010}. The temporal multimode capacity of the AFC scheme is linearly proportional to the number of teeth of the comb \cite{Afzelius2009a,Ortu2022}, which is one of the key benefits of the scheme.

To achieve on-demand read-out and longer storage times, one can map the optical \ket{g}--\ket{e} coherence into a \ket{g}--\ket{s} spin coherence through the application of a coherent population inversion pulse on the optical \ket{s}--\ket{e} transition \cite{Afzelius2009a}, see Fig. \ref{fig:AFC_scheme}a-b. The spin storage time $T_S$ is then limited by the inhomogeneous spin broadening \cite{Afzelius2010}, typically limiting the storage time in the range of 10 to 100 $\mu$s. Through the application of spin echo sequences on the \ket{g}--\ket{s} transition, one can further extend $T_S$ to the spin coherence time \cite{Jobez2015}, and potentially further through dynamical decoupling (DD) of the spin coherence from the dephasing environment  \cite{Holzaepfel2020, Ortu2021, Ma2021}. In practice, the spin echo sequence consists of a series of coherent population inversion pulses. The minimum sequence to perform the reversible back-and-forth mapping to the spin state, including the spin echo sequence, consists of two optical and two spin population inversion pulses \cite{Holzaepfel2020}, as shown in Fig. \ref{fig:AFC_scheme}b. The main goal of this article is to study and optimize this minimum sequence, and to show that it can approach unity in a practical experimental setting.

The total storage efficiency $\eta_{\rm tot}$ of the AFC spin-wave scheme depends on the purely optical AFC echo efficiency $\eta_{\rm AFC}$, the efficiency of each optical $\eta_{\rm OC}$ and spin $\eta_{\rm SC}$ population control pulse, and the spin dephasing $\eta_{\rm spin}$ during the duration $T_S$ \cite{Afzelius2010,Holzaepfel2020}:

\begin{equation}
\eta_{\rm tot}=\eta_{\rm AFC}(\eta_{\rm OC})^2(\eta_{\rm SC})^2\eta_{\rm spin}.
\label{eq:eta-tot}
\end{equation}

\noindent The factor $\eta_{\rm AFC}$ depends on the optical depth, shape and finesse, of the comb, as well as optical decoherence during the total time $1/\Delta$ spent in the excited state \ket{e}, see discussion in Ref. \cite{Ortu2022}. Experimental AFC efficiency studies have been focused on the optical $\eta_{\rm AFC}$ efficiency, reaching 35-41\% with the crystal in free space configurations \cite{Amari2010,Ortu2021}, and 53-62\% with the crystal in a cavity configuration \cite{Jobez2014,Sabooni2013,Duranti2024}. Of equal importance for spin-wave storage are the remaining factors in Eq. \ref{eq:eta-tot}, which have not been studied in detail. The topic of this paper is to measure, model and optimize the optical-to-spin conversion, not including spin dephasing, which we define by

\begin{equation}
\eta_{\rm opt-spin}=(\eta_{\rm OC})^2(\eta_{\rm SC})^2.
\label{eq:eta-opt-spin}
\end{equation}

\noindent Under the assumption that the spin dephasing is negligible ($\eta_{\rm spin} \approx 1$), for short enough spin storage times $T_S$, the $\eta_{\rm opt-spin}$ efficiency can be measured experimentally by recording the intensity of the output signal with and without applying the optical and spin control pulses, denoted $I^{\rm AFC-SW}_{\rm out}$ and $I^{\rm AFC}_{\rm out}$, respectively, and then compute the ratio

\begin{equation}
\eta_{\rm opt-spin}  \approx I^{\rm AFC-SW}_{\rm out}/I^{\rm AFC}_{\rm out}.
\label{eq:eta-rel}
\end{equation}

To reach a high $\eta_{\rm opt-spin}$ efficiency, one needs efficient and coherent population inversion pulses, both in the optical and spin domain. These must efficiently invert the entire AFC bandwidth and the inhomogeneous spin broadening. Adiabatic and frequency-chirped pulses can meet these requirements by properly time modulating the phase and amplitude of the field \cite{Hioe1985,Torosov2011}, and in principle they can reach unit efficiency for any bandwidth and available Rabi frequency \cite{Minar2010}, provided that they can be made long enough. Such pulses were developed in the NMR community, eg. the complex hyperbolic secant (sech) pulse \cite{Silver1985,Hioe1985}, and have also been applied for optical population transfer \cite{Rippe2005} and rephasing of nuclear spins \cite{Lauro2011,Mieth2012} for quantum memories. 

Here we will employ the ''extended'' sech pulse proposed by Tian et. al \cite{Tian2011}, called the hyperbolic-square-hyperbolic (HSH) pulse, which make particularly efficient use of the finite cut-off duration $T_C$ associated with the pulse, see Fig. \ref{fig:AFC_scheme}b. Technically, it has two smooth sech-type rising and falling edges, and a flat intensity part in the middle of duration $T_{sq}$. The frequency chirp ramps up and down smoothly at the edges, following the tanh function of a sech pulse, while being linear in the middle. In the adiabatic limit of a long pulse, the population inversion approaches unity over a flat bandwidth given only by the total chirp bandwidth $\Gamma_{\rm HSH}$. By applying two identically chirped pulses, one cancels the dynamic phases due to the chirp, which in principle can result in an identity operation on the initial coherence elements of the density matrix \cite{Minar2010}. In the numerical model and experiments presented in this article we will use two identical HSH pulses, both for the optical and the spin inversion sequence.

\section{Experimental implementation in \euyso{} with lifted Zeeman degeneracy}
\label{sec:exp_impl}

In this section, we will describe the main conceptual and experimental differences when implementing quantum memories under the application of a magnetic field in \euyso{}. The ideas and methods can be directly applied to the analogous \pryso{} system \cite{Gundogan2015, Rakonjac2021}. For a more general description of experimental details and methods in the context of AFC spin-wave memories, see Refs. eg. \cite{Laplane2016a,Jobez2016}.

\begin{figure}[h]
    \includegraphics[width=0.4\linewidth]{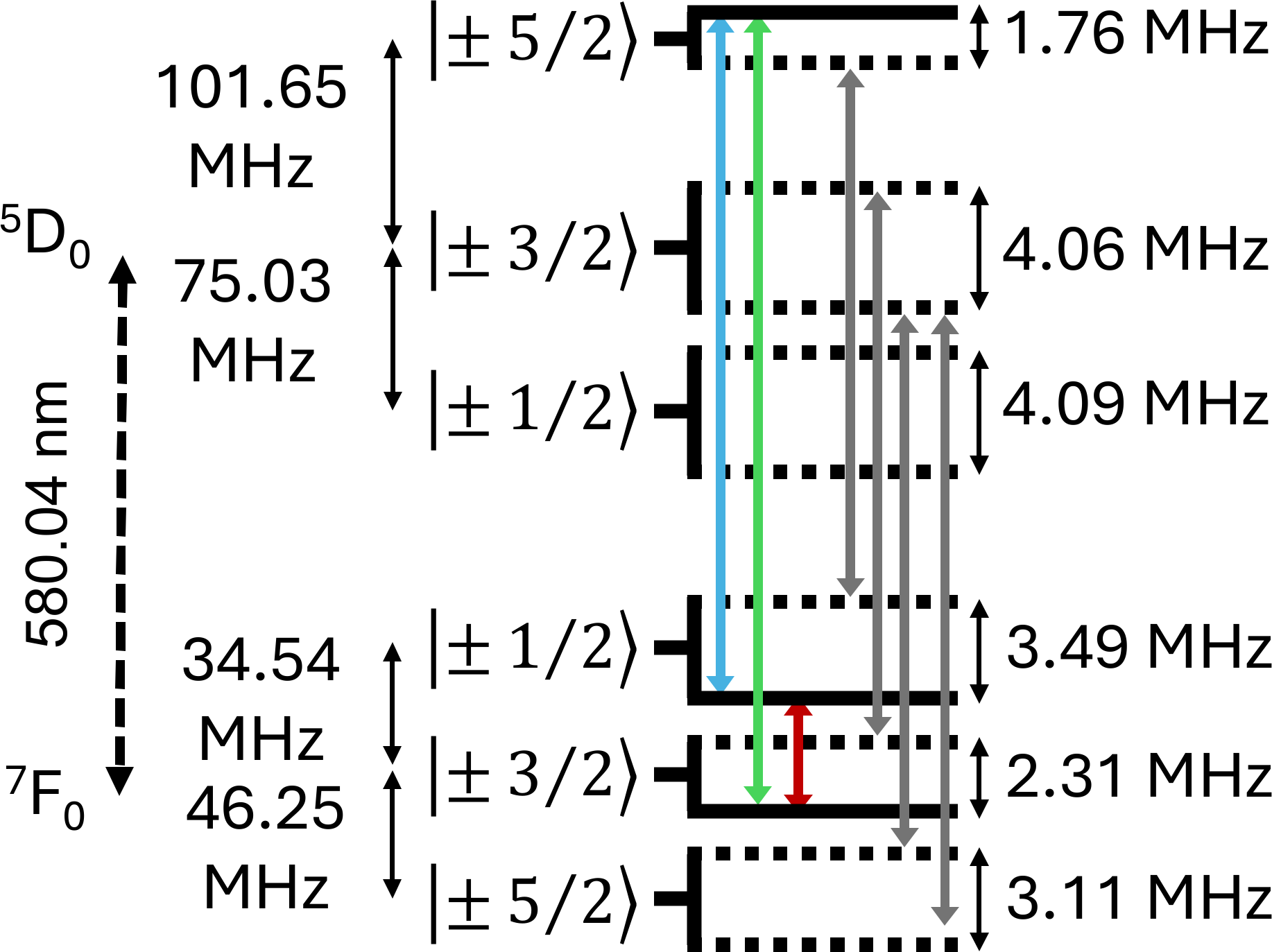}
    \centering
    \caption{The hyperfine levels of the electronic $^7$F$_0$ ground and $^5$D$_0$ excited states for site I of \euyso{} under the application of a magnetic field of $(231\pm4)$ mT at an angle of $(6.9\pm0.8)^{\circ}$ to the D$_2$ axis. The input/output (blue), optical (green) and spin (red) control transitions involved in the AFC spin-wave memory are shown, along with the optical pumping transitions (grey) employed for class cleaning and population initialization into the the \ket{g} state. Note that the energy separations are not to scale.}
    \label{fig:EuYSOlevels}
\end{figure}

Europium and praseodymium ions doped into \yso{} have quenched electronic spin, and the hyperfine structure is due to the $I=5/2$ nuclear spin and pure quadrupolar and pseudoquadrupolar interactions \cite{Longdell2002,Longdell2006}. As a result, at zero applied magnetic field, each crystal field level consists of three degenerate Zeeman doublets, denoted by \ket{\pm 1/2}, \ket{\pm 3/2} and \ket{\pm 5/2}. In the present work in \euyso{}, we consider the optical transition between the lowest-energy crystal field levels in the $^7$F$_0$ ground and the $^5$D$_0$ excited states, see Fig. \ref{fig:EuYSOlevels}. For the crystallographic site I in \euyso{}, this transition has the vacuum wavelength 580.04 nm \cite{Koenz2003}. The zero-field quadrupolar splits are 34.54 and 46.25 MHz for level $^7$F$_0$ \cite{Longdell2006,ZambriniCruzeiro2018a}, and 75.03 and 101.65 MHz \cite{Ma2018} for level $^5$D$_0$.

The bulk of quantum memory experiments in \euyso{} have employed the 34.54 and 46.25 MHz quardupole transitions at zero magnetic field for the \ket{g}--\ket{s} spin transition in the AFC scheme, or very close to zero field (a few mTesla) \cite{Jobez2015,Jobez2016,Laplane2016a,Laplane2017,Holzaepfel2020,Ortu2021,Holzaepfel2023}. Even at nominally zero applied field, the earth magnetic field and magnetic bias fields in the lab will cause a weak lifting of the degeneracy, as well as defining the relative transition moments of both the optical and spin transitions between all the hyperfine levels. This in turn will affect the optical and spin manipulation through the population inversion pulses \cite{Etesse2021}. Moreover, the weak splits of the Zeeman doublets will cause time-dependent modulation of the AFC memory efficiency \cite{Etesse2021}, due to interference of different coherence elements in the multi-level density matrix. In such a scenario, it is unclear if the optical-to-spin conversion can be made perfectly reversible.  To avoid such effects, we here work in the regime of a moderately strong magnetic field of around 230 mT, allowing us to split the Zeeman doublets such that all optical and spin transitions involved in the AFC spin-wave scheme are spectrally resolvable. The \ket{g}, \ket{e} and \ket{s} states in the scheme are thus single magnetic sublevels, neglecting effects such as superhyperfine interactions \cite{Fraval2004a,Pignol2024}.

The effective spin Hamiltonian of the ground and excited levels are known, both for \euyso{} \cite{Longdell2006,ZambriniCruzeiro2018a,Ma2018} and \pryso{} \cite{Longdell2002}, allowing calculations of the hyperfine energies, as well as optical transition moments connecting different hyperfine states, for an arbitrary magnetic field orientation. There are two magnetic subsites for each crystallographic site in \yso{}, which are related by a C$_2$ symmetry along the crystallographic b-axis. Here, we work in the D$_1$-D$_2$ plane \cite{Li1992} perpendicular to the b axis, where the two magnetic subsites are equivalent, such that all \eu{} ions have the same energy structure. Another important consideration is the bandwidth over which a single class of ions can be prepared with the usual methods applied in rare-earth-ion crystals, whose general concepts are described in detail elsewhere \cite{Nilsson2004,Lauritzen2012}. At close to zero magnetic field in \euyso{}, a single class of ions can be prepared over 5 MHz bandwidth \cite{Ortu2022}, limited by the separation of some specific optical-hyperfine transitions. To gain insight into the limitations of the bandwidth when applying a magnetic field, simulations of the SHB sequence were carried out for different angles in the D$_1$-D$_2$ plane. From these simulations, we conclude that the largest bandwidth is reached with the field close to the D$_2$ axis, with the bandwidth being limited by the split of the excited state \ket{\pm 5/2} doublet. The simulation results will be presented in a separate article \cite{Chen2024}.

Experimentally, a moderate magnetic field of about 230 mT was applied by a set of permanent NdFeB magnets in a sandwich configuration around the crystal. The crystal had a length of 12.5 mm along the light propagation direction (b axis), and the magnets were 20 mm long to reduce field inhomogeneities. The magnets were solidly mounted inside the cryocooler vacuum chamber, with its field direction roughly aligned in the crystal D$_2$ axis. The magnets and the \euyso{} crystal were cooled down to about 3.3 K. It is critical to align the field in the D$_1$-D$_2$ plane, to avoid a split of all the optical and spin transitions due to non-equivalent magnetic subsites. To this end, we applied a corrective magnetic field along the b axis, using large coils external to the cryocooler. The field was accurately fine tuned by measuring both optical SHB lines and spin resonances, using Raman heterodyne scattering (RHS) \cite{Mlynek1983}, and observing the merging of the transition frequencies due to the two magnetic subsites, as the the field was tilted into the D$_1$-D$_2$ plane. The corrective field magnitude was generally in the range of 8-10 mT. Once the merging ''cross point'' was found, the magnetic field vector $\mathbf{B}$ was estimated by fitting a set of measured SHB frequencies to the predictions from the spin Hamiltonian, allowing a precise estimation of both the field magnitude and the angle in the D$_1$-D$_2$ plane. Experiments presented in this article were performed with an estimated field magnitude of $(231\pm4)$ mT at an angle of $(6.9\pm0.8)^{\circ}$ to the D$_2$ axis of the crystal. The calculated Zeeman splits for this field are shown in Fig. \ref{fig:EuYSOlevels}. Note that the excited state \ket{\pm 5/2} doublet split is 1.76 MHz, which limits the bandwidth over which a single class can be prepared. The AFC spin-wave transitions and the auxiliary transitions employed in the class cleaning and spin polarization steps, are also shown in Fig. \ref{fig:EuYSOlevels}. Note that in total six optical transitions are involved, one for each ground state, in contrast to the three transitions employed in previous AFC spin-wave experiments close to zero field \cite{Jobez2015,Laplane2016a,Ortu2021}. The input/output transition is a strong transition, while the control pulse transition is moderately strong, see \ref{app:branching-table}. The auxiliary transitions were chosen by considering both transition strengths and their total frequency spread with respect to the acousto-optic modulator employed for the frequency tuning. More details about the experimental set-up, the SHB simulations, magnetic field alignement and estimation methods, as well as measurements of some relative transition strengths, are presented in an associated article \cite{Chen2024}.

\section{Numerical modelling of the optical-to-spin efficiency}
\label{sec:num-simul}

To understand how different experimental parameters influence the optical-to-spin conversion efficiency, we use numerical modelling to compute the efficiencies of the optical and spin population inversion pulses. A key factor in this context is the Gaussian intensity distribution of both the input/output mode (named input mode in the following) and the optical control mode, which is often neglected. The goal of the model is to be able to compute the total efficiency, including the spatial distribution and overlap of these modes as they propagate through the crystal, possibly at an angle $\theta$. The model is based on a spatial average of the transfer efficiency across and along the propagation direction. To simulate the coherent interaction between the optical and spin pulses with the atoms at any spatial point along the propagation direction, we employ Bloch equations in order to calculate the frequency-averaged population inversion efficiency over the relevant bandwidth, both in the optical and spin domain. The main features and assumptions of the numerical model will be described briefly here. We will start by describing the radial model when the input/output mode and the control mode are co-propagating, $\theta = 0$, and then briefly describe its extension to the case of a non-zero $\theta$ between the modes.

\subsection{Spatial efficiency averaging}
\label{sec:num-simul-sa}

In both the optical and spin domain, one can imagine spatially varying fields in experiments. However, for our experimental implementation, the spatial variation of the optical fields is much more severe, due to their Gaussian nature, while the RF field produced by the coil used for spin manipulation will have much less of spatial variation. In addition, any RF inhomogeneity is considerably harder to model, while in the optical domain we can rely on well-behaved Gaussian modes. We will thus only consider the spatial variation of the optical control pulses.

The optical control mode is assumed to have its waist $w_0^c$ \footnote{We use the standard definition of the waist radius $w_0$ at the level $\exp(-1)$ of the field amplitude.} placed at $z = 0$ in the middle of the crystal of length $L$, i.e. $z \in [-L/2,L/2]$, with a Rayleigh length $z_R^c$. A key input parameter of the model is the peak Rabi frequency in the middle of the crystal, $\Omega_c^0$, which must be estimated from experiments or calculated from the peak power and the transition dipole moment. The input mode is defined similarly, with a waist $w_0^{in}$ at $z=0$ and a Rayleigh length $z_R^{in}$. For the calculation of an inversion pulse efficiency, we introduce the radially symmetric intensity distribution $I(r,z)$ of the input mode, which is area normalized at each $z$ such that $2 \pi \int_{0}^{\infty} I(r,z) r dr = 1$. 

To reach a high total efficiency spatially averaged over $r$ and $z$, it is clear that the input mode should have a smaller waist radius, and hence a shorter Rayleigh length, as compared to the control mode. Yet, the control mode should be as tight as possible, to maximize the peak Rabi frequency $\Omega_c^0$. It is thus important to minimize the volume occupied by the input mode, over the entire crystal length. We here introduce the input mode waist $w_{0,opt}^{in}$ which minimizes the input mode radius at the surface of the crystal at $z = \pm L/2$. This minimum is obtained when setting the input mode Rayleigh length to $z_R^{in} = L/2$, such that 

%optimal input waist formula
\begin{equation}
w_{0,opt}^{in} = \sqrt{\frac{L}{2}\frac{\lambda}{\pi}},
\label{eq:opt-waist}
\end{equation}

\noindent where $\lambda$ is the wavelength. Intuitively, this choice minimizes the volume occupied by the input waist, allowing a tighter control waist.

The Gaussian distribution of the control mode Rabi frequency $\Omega_c(r,z)$, with both radial $r$ and propagation $z$ dependence, results in a spatially varying population inversion efficiency $\eta_{\bar{\delta}}(r,z) \leq 1$, for each single optical control pulse. Note that by its definition $\Omega_c^0 = \Omega_c(0,0)$. Importantly, $\eta_{\bar{\delta}}(r,z)$ represents a spectrally averaged efficiency in the detuning space $\delta$, taking account the relevant spectral distribution in the optical/spin domain, see Sec. \ref{sec:num-simul-da}. The efficiency at each distance $z$ is calculated through the radial integration $2 \pi \int_{0}^{\infty} I(r,z) \eta_{\bar{\delta}}(r,z) r dr$, and the total single-pulse efficiency $\eta_{\rm OC}^{\rm th}$ is calculated by a spatial $z$-average, defined as

\begin{equation}
\eta_{\rm OC}^{\rm th} = \frac{2\pi}{L} \int_{-L/2}^{L/2} \int_{0}^{\infty} I(r,z) \eta_{\bar{\delta}}^{\rm opt}(r,z) r dr dz,
\label{eq:eta-spatial-int-OC}
\end{equation}

\noindent where $\eta_{\bar{\delta}}^{\rm opt}(r,z)$ denotes the spectrally averaged efficiency over the spectrum in the optical domain.

For the spin control pulses, we assume a spatially homogeneous field, such that its single-pulse efficiency $\eta_{\rm OC}^{\rm th}$ is simply

\begin{equation}
\eta_{\rm SC}^{\rm th} = \eta_{\bar{\delta}}^{\rm spin},
\label{eq:eta-spatial-int-SC}
\end{equation}

\noindent where $\eta_{\bar{\delta}}^{\rm spin}$ denotes the spectrally averaged efficiency over the spectrum in the spin domain.

The total numerical model efficiency, including two optical and two spin control pulses, is thus given by

%theoretical total optical-spin efficiency
\begin{equation}
\eta_{\rm opt-spin}^{\rm th}=(\eta_{\rm OC}^{\rm th})^2(\eta_{\rm SC}^{\rm th})^2.
\label{eq:eta-opt-spin-model}
\end{equation}

\subsection{Spectral efficiency averaging}
\label{sec:num-simul-da}

To calculate the spectrally averaged inversion efficiency $\eta_{\bar{\delta}}(r,z)$ for each spatial position, we employ standard two-level Bloch equations for simulating the light matter interaction \cite{Boyd1992}. Briefly, the Bloch vector $\vec{r}(\delta)$ is calculated numerically for each spatial position, given the control mode Rabi frequency $\Omega_c(r,z)$, as a function of the detuning $\delta$. The model assumes no population decay, $T_1 = \infty$, nor any pure dephasing, $T_2 = \infty$, although these can be easily included if necessary. In both the optical and spin domain, the efficiency should be averaged over the relevant spectrum. In the optical domain, we assume this to be the spectrum of the optical pulse in the input mode, which is approximately Gaussian in our case. Here we assume that the AFC bandwidth is larger than both the input spectrum and the optical HSH bandwidth, such that the AFC bandwidth is not a model parameter. The model can handle, however, any relation between the input spectrum and the optical HSH control pulse bandwidth. In the spin domain, we take the average over the entire inhomogeneous spin linewidth. For both domains, we assume a Gaussian distribution $S(\delta)$ with a full-width at half-maximum linewidth of $\Gamma_{\rm opt}$ and $\Gamma_{\rm spin}$, respectively, where $S(\delta)$ is area-normalized $\int_{-\infty}^{\infty} S(\delta) d\delta = 1$. From the simulated Bloch vector $\vec{r}(\delta) = (r_1(\delta),r_2(\delta),r_3(\delta))$, the population transfer efficiency is calculated through the population inversion component $r_3(\delta)$, the efficiency being $(1+r_3(\delta))/2$. The final efficiency $\eta_{\bar{\delta}}(r,z)$, averaged over the spectrum, is thus

%Bloch detuning-averaged efficiency
\begin{equation}
\eta_{\bar{\delta}}(r,z) = \frac{1}{2} \int_{-\infty}^{\infty} S(\delta) (1+r_3(\delta))  d\delta .
\label{eq:eta-Bloch}
\end{equation}

The Bloch model can take any type of population inversion pulse as an input, provided that the maximum Rabi frequency in the time domain is set to $\Omega_c^0$.  Here, we will restrict ourselves to the HSH pulse introduced in Sec. \ref{sec:exp_impl}, whose time-domain phase and amplitude modulation is described in detail in Ref. \cite{Tian2011}. In the majority of simulations presented in the paper, we use a full numerical calculation using the Bloch equations, allowing us to calculate the model for any type of HSH pulse. In some cases, we replace the numerical calculation of the optical HSH pulses with an analytical one. For a long adiabatic HSH pulse, where the frequency spectrum is entirely given by the frequency chirp of the pulse, $\Gamma_{\rm HSH} \gg 1/T_{sq}$, then the pulse efficiency can be calculated analytically by (see Supplemental Material of Ref. \cite{Businger2020}), 

\begin{equation}
	\eta_{\bar{\delta}}(r,z) = 1-\exp\left( -0.5 \pi \frac{T_{sq} \Omega_c(r,z)^2}{\Gamma_{\rm HSH}} \right).
	\label{eq:HSH_eff}
\end{equation}

\noindent The unit of all frequency parameters are in rad/s. This formula assumes implicitly that the spectrum $S(\delta)$ is entirely covered by the HSH chirp range $\Gamma_{\rm HSH}$. The analytical formula provides a very large speed-up of the calculations, but it is not accurate for shorter HSH pulses, which can still be efficient and of interest in this context. In the following sections, we will compare the full Bloch model and the simplified analytical model for the optical control pulses. For the spin HSH pulses we always use the numerical model, as the function is only used sparsely in the model.

%crossed-beam model
\subsection{Crossed-beam configuration}
\label{sec:num-simul-cbc}

The extension of the model to a non-zero angle $\theta$ between the optical modes is straightforward. The  $\eta_{\bar{\delta}}(x,y,z)$ efficiency must now be calculated as a 2-d matrix, with the Cartesian coordinates $x$ and $y$ in the plane perpendicular to $z$. However, the Gaussian control mode is still radially symmetric around its centre, and its symmetric efficiency $\eta_{\bar{\delta}}(r,z)$ can still be calculated efficiently using the radial model, which speeds up the calculation. If we assume that the input and output modes propagate with an angle in the $x$ axis, then the Cartesian efficiency function $\eta_{\bar{\delta}}(x,y,z)$ can simply be computed from the radial one $\eta_{\bar{\delta}}(r,z)$, with a displacement $\Delta  x = z \tan(\theta)$ along the $x$ coordinate. The pulse efficiency is calculated through $x$,$y$ and $z$ averaging, in analogy to Eq. \ref{eq:eta-spatial-int-OC} for the co-linear case.

\begin{figure}[h]
    \centering
	\begin{subfigure}{0.45\textwidth}
     		\centering
	   	\includegraphics[scale=0.55]{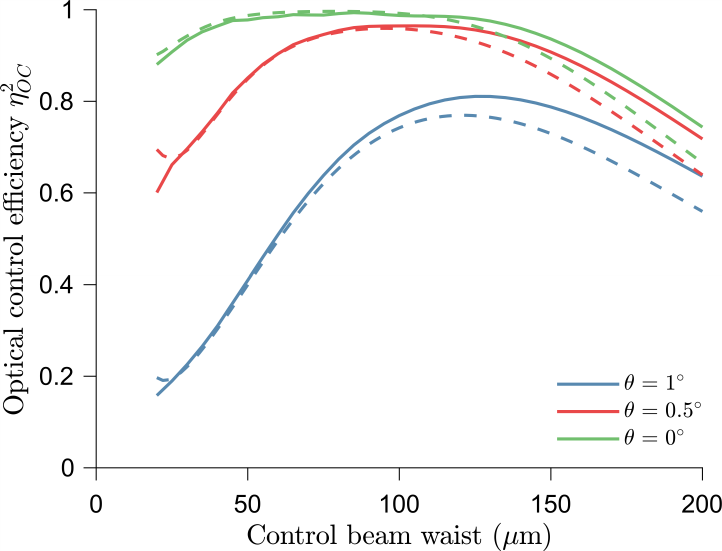}
     \end{subfigure}        
	\begin{subfigure}{0.45\textwidth}
     		\centering
     		\includegraphics[scale=0.55]{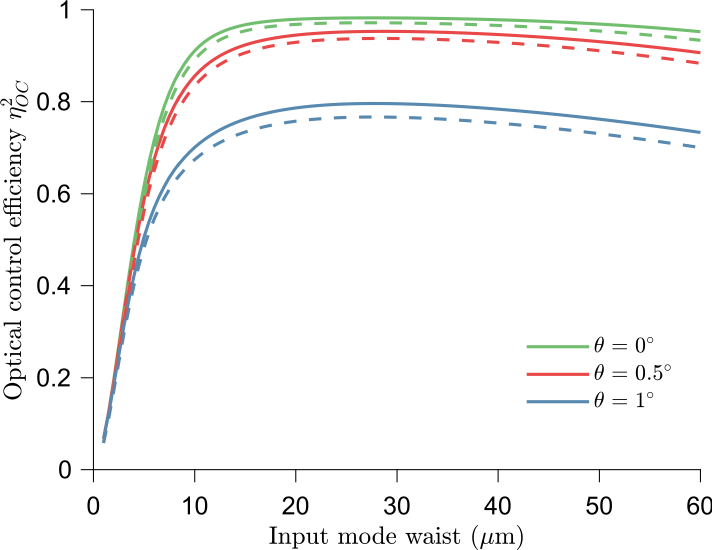}
     \end{subfigure}        
   \caption{(a) Numerically calculated total optical control efficiency $(\eta_{\rm OC}^{\rm th})^2$, as a function of the control waist $w_0^c$, for the angles $\theta$ = 0 (green), 0.5 (red) and 1 (blue) degrees. Solid lines represent the Bloch model, the dashed lines the analytical HSH model, see Sec. \ref{sec:num-simul-sa}. The input mode waist radius was $w_0^{in} = 34$ $\mu$m and the control Rabi frequency was set to the experimental value of $\Omega_c^0/(2 \pi) = 532$ kHz for $w_0^c = 60$ $\mu$m. For other control waists in the plot, $\Omega_c^0$ was scaled with $w_0^c$ by $\Omega_c^0 \propto 1/w_0^c$. The optical HSH pulse had a total duration of $T_{C} = 15$ $\mu$s, and the remaining input parameters are given in Sec. \ref{sec:exp-params}. (b) $(\eta_{\rm OC}^{\rm th})^2$ as a function of the input mode size, for a control beam waist of 120 $\mu$m. All other parameters at the same as in panel (a).}
	\label{fig:model-ctrl-waist-scan}
\end{figure}

To illustrate the full 2-d Bloch model, in Fig. \ref{fig:model-ctrl-waist-scan}(a) the total optical control pulse efficiency $(\eta_{\rm OC}^{\rm th})^2$ is shown as a function of the control waist $w_0^c$, for a given input waist and three different angles $\theta$. The input parameters correspond to experimentally measured values in this work, detailed in Sec. \ref{sec:exp-params}. These simulations show that the optimum control mode waist increases with the angle, as expected. The highest efficiency also decreases inevitably, due to the finite Rabi frequency and the fixed HSH pulse duration. In \ref{app:mode_profiles}, examples of the transverse mode profiles of the input, control and the pulse efficiency are shown. 

While the choice of the control mode is important for larger angle, the efficiency is surprisingly constant when varying the input mode waist, for a given control mode waist, as shown in Fig. \ref{fig:model-ctrl-waist-scan}b. This is presumably due to the fact that the control efficiency depends more strongly on the waist, as the Rabi frequency scales inversely with the waist. The input mode given by Eq. \ref{eq:opt-waist} achieves close to the maximum efficiency, but clearly the input waist choice is not critical.

\section{Experimental results}
\label{sec:exp-results}

\subsection{Experimental parameters}
\label{sec:exp-params}

The experiments presented throughout the article were performed with a common set of parameters given in this section. In the following sections, we will present experiments where one of the parameters was varied with respect to the nominal value given in this section, other parameters were identical, unless otherwise stated.

The input mode was focused in the crystal with a waist radius $w_0^{in} = 34$ $\mu$m, calculated using Eq. \ref{eq:opt-waist}, for the transition wavelength $\lambda = 580$ nm and crystal length $L = 12.5$ mm. Another spatial mode was used for the class cleaning, spin polarization, and AFC creation pulses, with a large mode waist of 350 $\mu$m, to achieve a spatially homogeneous preparation. The single-pass optical depth on the input transition \ket{-1/2g} $\leftrightarrow$ \ket{+5/2e} transition, measured through the spatial input mode, was $d = 2.4$ after the spin polarization step, which resulted in a short-delay AFC efficiency of $\eta_{\rm AFC} = (17.9 \pm 0.2)$\% at $1/\Delta = 3 \mu$s. All the following measurements were done at a longer delay of $1/\Delta = 30 \mu$s, to create ample space for the optical control pulses, resulting in a lower AFC efficiency of $\eta_{\rm AFC} = (7.4 \pm 0.2)$\%, see \ref{app:AFC_echo}.

The optical control field was applied on the \ket{-3/2g} $\leftrightarrow$ \ket{+5/2e} transition. For an optical peak power of 385 mW, we obtained Rabi frequencies of $\Omega_c^0/(2 \pi)$ = $(263 \pm 5)$, $(336 \pm 4)$ and $(532 \pm 11)$ kHz, for waists radii of $w_0^c$ = 120, 100 and 60 $\mu$m, respectively. The co-linear experiments used the tighter control mode of $60 \mu$m, while the crossed-beam experiments employed the larger mode of $120 \mu$m. 

The Gaussian input mode FWHM intensity duration was set to $4\mu$s, corresponding to a FWHM power spectrum of $\Gamma_{\rm opt}/(2 \pi)  = 110$ kHz. The optical HSH pulses had a square part $T_{sq} = 0.7 T_C$ and a total duration of $T_C = 15~\mu$s, unless stated differently, and the control pulse duration hereafter refers to the total cut-off duration $T_C$. The chirp bandwidth was set to $\Gamma_{\rm HSH}/(2 \pi) = 1.5$ MHz, slightly less than the experimental maximum bandwidth set by the Zeeman splitting, as discussed in Sec. \ref{sec:exp_impl}, but significantly larger than the optical input bandwidth of 110 kHz. This allows to probe the highest possible achievable $\eta_{\rm opt-spin}$, without an accurate measurement of the input pulse spectrum. 

The spin control excited the \ket{-3/2g} $\leftrightarrow$ \ket{-1/2g} hyperfine transition, at 34.00 MHz, reaching a Rabi frequency of $\Omega_{\textrm{rf}}^0/(2 \pi)$= 42 kHz at the full RF power of about 100 W. The spin HSH pulses had a square part $T_{sq} = 0.75 T_C$, with a cut-off duration of $T_C = 200 \mu$s, with a chirp bandwidth of $\Gamma_{\rm HSH}/(2 \pi)$ = 350 kHz, unless otherwise stated. The inhomogeneous spin linewidth was measured to be $\Gamma_{\rm spin}/(2 \pi)  = 123$ kHz. The nominal spin storage time was $T_S$ = 500 $\mu$s.

\subsection{Optimizing the spin sequence}

We start by discussing the optimization of the spin control sequence, for the co-linear optical configuration, $\theta = 0$ deg, and the control waist of 60 $\mu$m. A key parameter of the spin sequence is the HSH pulse bandwidth $\Gamma_{\rm HSH}$ of the RF spin control pulses. It should be large enough to efficiently invert the entire spin broadening. However, a too large bandwidth will reduce the efficiency, for a fixed Rabi frequency and pulse duration, see Eq. \ref{eq:HSH_eff}. In addition, to enhance the RF Rabi frequency, we employ a resonant RF circuit with the coil \cite{Ortu2021a}, impedanced matched to the RF amplifier at 50 Ohm. The circuit introduces a frequency-dependent Rabi frequency, which we model with a Lorentzian function, such that

\begin{equation}
	\Omega_{\textrm{rf}}(\omega) = \Omega_{\textrm{rf}}^0 \frac{(\Gamma_{\textrm{rf}}/2)^2}{(\omega-\omega_0)^2 + (\Gamma_{\textrm{rf}}/2)^2},
	\label{eq:RF_Rabi}
\end{equation}

\noindent  where $\Gamma_{\textrm{rf}}$ is the circuit FWHM and $\omega_0$ it's resonance frequency tuned as close as possible to 34.0 MHz.

\begin{figure}[h!]
    \centering
	\begin{subfigure}{0.45\textwidth}
     		\centering
	   	\includegraphics[scale=0.55]{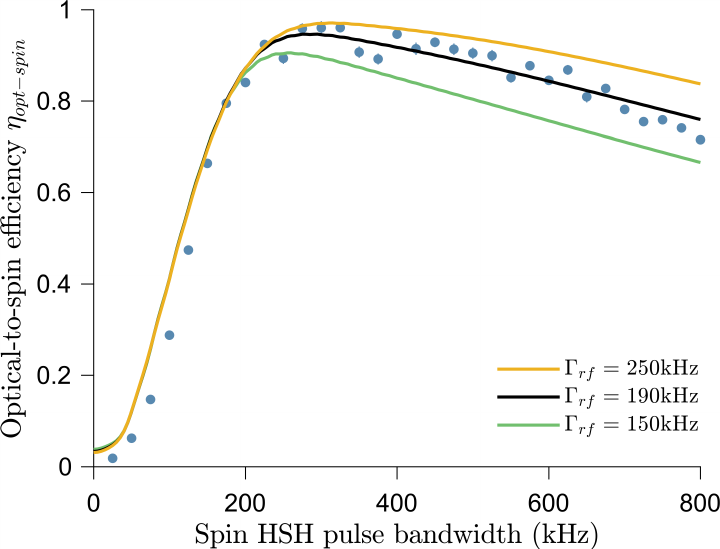}
     \end{subfigure}        
	\begin{subfigure}{0.45\textwidth}
     		\centering
		\includegraphics[scale=0.55]{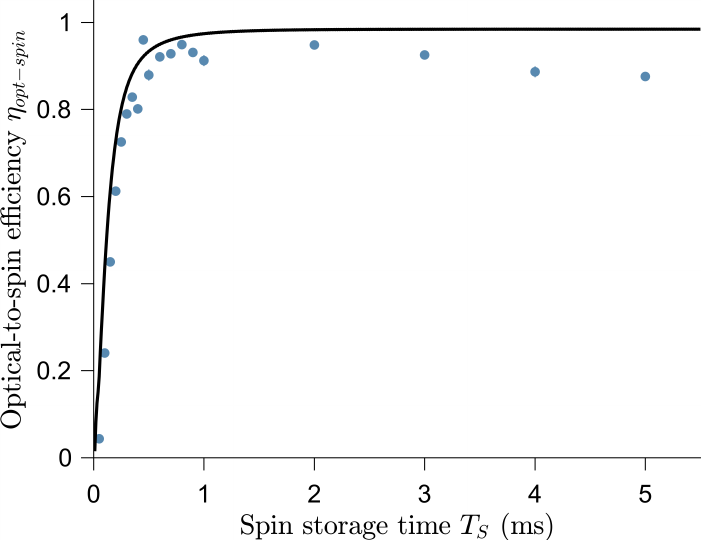}
     \end{subfigure}        
    \caption{(a) The experimental optical-to-spin efficiency $\eta_{\rm opt-spin}$ for the co-linear beam configuration, $\theta = 0$ deg, measured as a function of the HSH chirp bandwidth of the spin (RF) population inversion pulses. The solid lines represent the radial theoretical model, for optimized RF circuit linewidth of 190 kHz (black), and for two other cases of 150 kHz (yellow) and 250 kHz (green) linewidths. (b) The efficiency $\eta_{\rm opt-spin}$ measured as a function of the total spin storage time $T_S$, compared to the model (solid black line) based on the optimized circuit linewidth. See text for more details regarding the measurements and modelling.}
    \label{fig:RF_opt}
\end{figure}

Figure \ref{fig:RF_opt}(a) shows the measured optical-to-spin efficiency $\eta_{\rm opt-spin}$, Eq. \ref{eq:eta-rel}, for the co-linear beam configuration, as a function of the HSH bandwidth of the RF pulses. The efficiency first increases as more of the 123 kHz spin linewidth is covered, then starts to drop primarily due to the circuit linewidth. A maximum efficiency of about $(96\pm1)\%$ is reached in this measurement, for a HSH bandwidth of about 300 kHz. In the following measurements, the bandwidth was set to a slightly higher value of $350$ kHz, in order to be less sensitive to possible variations in the circuit response function.

The radial Bloch model discussed in Sec. \ref{sec:num-simul} accurately predicts the efficiency as a function of RF HSH bandwidth, including the maximum attainable efficiency, provided that we adjust the RF circuit bandwidth. We emphasize that the model includes all the relevant parameters of both the optical and spin control pulses, including the spatial field distributions of the optical pulses, as discussed in detail in Sec. \ref{sec:num-simul}. The best fit is obtained with $\Gamma_{\textrm{rf}}/(2 \pi) = 190$ kHz, which will be used for all other simulations presented in the paper.

In Figure \ref{fig:RF_opt}(b), the spin storage time was varied with respect to its nominal value of $T_S$ = 500 $\mu$s, between 50 $\mu$s and 5 ms. For storage times below 500 $\mu$s, the spin HSH cut-off duration $T_C$ had to be reduced, and it was set to 80\% of $T_S/2$, while for above 500 $\mu$s it was fixed to its nominal value of $T_C = 200~\mu$s. For short storage times, the efficiency of the HSH pulse drops due to the shorter pulse duration $T_C$, cf. Eq. \ref{eq:HSH_eff}, which is well predicted by the radial Bloch model. For longer storage times, there is a weak decay stemming from the finite spin dephasing at the scale of 10s of ms. Its impact on the efficiency at $T_S$ = 500 $\mu$s is mostly negligable, as the model accurately predicts the maximum efficiency of about $96\%$, without including the spin dephasing. This justifies our use of Eq. \ref{eq:eta-rel} for calculating the optical-to-spin efficiency.

\subsection{Optical control pulse duration}

In this section, we study the optical-to-spin efficiency as a function of the optical control pulse duration, for the co-linear configuration, $\theta = 0$ deg, and the 60 $\mu$m control waist. The spin control sequence was fixed to the optimized sequence of the previous section.

\begin{figure}[h]
    \includegraphics[scale=0.55]{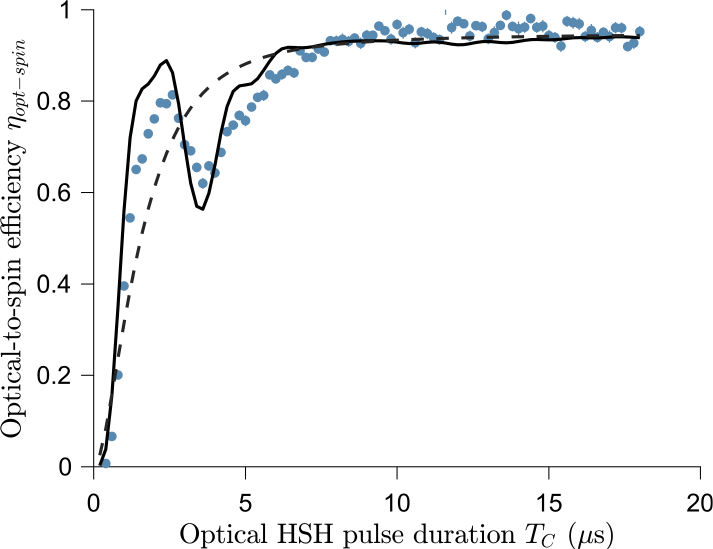}
    \centering
    \caption{The experimental optical-to-spin efficiency $\eta_{\rm opt-spin}$ for the co-linear beam configuration, $\theta = 0$ deg, measured as a function of the optical HSH cut-off duration $T_C$. The solid black line represents the radial model based on the Bloch simulation described in Sec. \ref{sec:num-simul}, using all relevant experimental parameters. The dashed line represents the same model, except that the spectral averaging of the optical pulse is calculated using the analytical HSH formula, Eq. \ref{eq:HSH_eff}.}
    \label{fig:ctrl_dur_scan}
\end{figure}

In Figure \ref{fig:ctrl_dur_scan}, we present the measured $\eta_{opt-spin}$ efficiency, as a function of the optical HSH cut-off duration $T_C$. As expected for an adiabatic pulse, for sufficiently long optical pulse durations the efficiency saturates. By taking all the points for durations of 10 $\mu$s and beyond, we calculate a mean efficiency of $\eta_{opt-spin} =95.5 $\% with a standard deviation of 1.8\%. For short pulse durations, when the Fourier spectrum of the pulse $1/T_{sq}$ is not negligible with respect to the RF HSH bandwidth $\Gamma_{\rm HSH}$, the adiabaticity of the pulse is not completely fulfilled, leading to low efficiencies. In the intermediate regime, the pulse efficiency oscillates weakly, which is well-explained by the radial Bloch model, as shown in Figure \ref{fig:ctrl_dur_scan}. The radial model based on the analytical HSH formula, Eq. \ref{eq:HSH_eff}, works well for long pulse durations, but is unable to predict the oscillations for short pulse durations.

\subsection{Input pulse detuning dependence}

\begin{figure}[h]
    \includegraphics[scale=0.55]{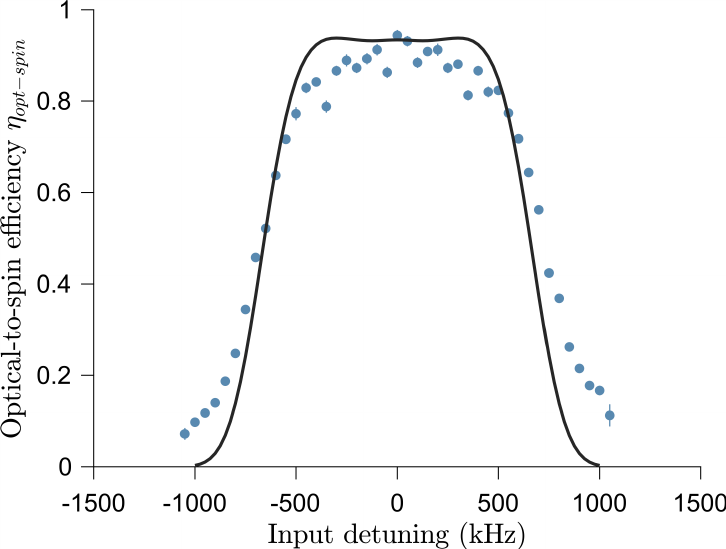}
    \centering
    \caption{The experimental optical-to-spin efficiency $\eta_{\rm opt-spin}$ for the co-linear beam configuration, $\theta = 0$ deg, as a function of the input pulse detuning. All other experimental parameters were identical to the experiment shown in Fig. \ref{fig:ctrl_dur_scan}.}
    \label{fig:input_det}
\end{figure}

An adiabatic control pulse can achieve high transfer efficiency over its chirp bandwidth, a main motivation for their use in AFC memories. In this work, the optical HSH pulses had a bandwidth of $\Gamma_{\rm HSH}/(2 \pi) = 1.5$ MHz. In Fig. \ref{fig:input_det}, we present measurements of the $\eta_{\rm opt-spin}$ efficiency, as a function of the detuning of the narrowband input pulse with respect to the centre of the optical HSH pulse, which in itself is centered on the AFC. The effective FWHM bandwidth of the optical-to-spin manipulation is 1.33 MHz, consistent with the chirp bandwidth. The radial Bloch model also accurately predicts the effective bandwidth. Note that the maximum efficiency is slightly lower than those observed in the other measurements, probably due to a slightly misaligned input-control mode overlap.

\subsection{Crossed-beam angle dependence}

Quantum storage experiments with input modes at the single photon level typically require spatial filtering, to reduce photon noise from the optical control fields. This is typically achieved using a crossed-beam configuration, with an angle $\theta$ between the input and control modes. For a cross-beam configuration, the choice of the control waist radius is more critical, in order to achieve a good mode overlap over the entire input mode in the crystal. From the full 2-d Bloch simulation shown in Fig. \ref{fig:model-ctrl-waist-scan}(a), it is clear that a larger control waist is optimal, with respect to the co-linear configuration. Based on the simulation, we set the control radius to 120 $\mu$m.

\begin{figure}[h]
    \includegraphics[scale=0.55]{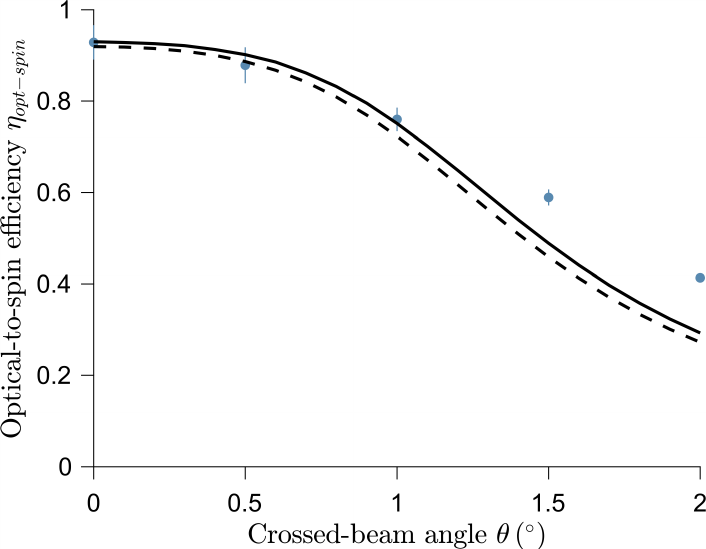}
    \centering
    \caption{The experimental optical-to-spin efficiency $\eta_{\rm opt-spin}$ for the crossed-beam configuration, as a function of the angle $\theta$ between the input and control modes. The solid line shows the simulation using the full 2-d Bloch model, and the dashed line the same simulation with the analytical HSH formula.}
    \label{fig:angle_scan}
\end{figure}

In Figure \ref{fig:angle_scan}, the measured optical-to-spin efficiency is shown for a range of angles, with a fixed optical HSH control duration of $T_C = 15~\mu$s and the optimized spin control sequence. As expected, the efficiency decreases with increasing angle, but remains larger than 75\% for angles up to $\theta = 1$ deg. The full 2-d Bloch model predicts an efficiency which is lower than the experimental values for the largest angles, which is not fully understood. Simulations with reasonable variations of the model parameters for the input waist, control waist and Rabi frequency cannot explain the discrepancy. It should be noted that the control mode waist should in principle be optimized for each angle, which can be done with the simulation tool, as shown in Fig. \ref{fig:model-ctrl-waist-scan}. For the angle $\theta = 2$ deg, the optimal control waist is about 160 $\mu$m, which would result in about 5\% higher efficiency according to simulations.

\section{Discussion and outlook}

In this article, we have focused on the modeling, optimization and measurement of the relative optical and spin manipulation  efficiency of an AFC spin-wave quantum memory, reaching an unparalled efficiency of $\eta_{\rm opt-spin} = 95-96\%$, including spin dephasing. For applications, the total memory efficiency $\eta_{\rm tot}$, including the optical AFC echo efficiency $\eta_{\rm AFC}$, see Eq. \ref{eq:eta-tot}, is the key parameter. In the current work, we obtained $\eta_{\rm AFC} = 7.4\%$ at the delay of $1/\Delta = 30 \mu$s, giving a total memory efficiency of $\eta_{\rm tot} = 7.1\%$. In principle, to optimize the total efficiency, one should consider that $\eta_{\rm AFC}$ increases as the $1/\Delta$ delay is shortened, see \ref{app:AFC_echo}, while the $\eta_{\rm opt-spin}$ efficiency decreases due to the need of using shorter optical control pulses, see Fig. \ref{fig:ctrl_dur_scan}. In addition, for short control pulses, it might be advantageous to re-optimize the control waist radius, as shown in Fig. \ref{fig:model-ctrl-waist-scan} for the 15 $\mu$s pulse. Such optimizations are time consuming experimentally, but the simulation tools developed in this work can quickly point to optimal configurations.

To significantly increase $\eta_{\rm AFC}$ requires larger optical depth of the crystal. A straightforward solution is to use a multi-pass configuration through the crystal, in a crossed-beam configuration. Up to four passes can be readily achieved, see for instance Ref. \cite{Businger2020}, using a small angle compatible with this work, which considerably increases the $\eta_{\rm AFC}$ efficiency. As an example, for the same AFC delay of $1/\Delta = 30~\mu$s, an efficiency of $\eta_{\rm AFC} = 24.8\%$ was obtained in double pass configuration in \euyso{} \cite{Ortu2022}. 

To reach still higher $\eta_{\rm AFC}$ efficiency requires a cavity approach \cite{Jobez2014,Sabooni2013,Duranti2024}, which has also been implemented in a crossed-beam configuration for spin-wave AFC memories \cite{Jobez2014}. For cavity crossed-beam experiments, it is highly interesting to consider short crystals of a few mm, to increase the efficiency for larger angles, provided that the lower absorption of the crystal can still be impedance-matched to the cavity \cite{Afzelius2010a}. In Figure \ref{fig:model-ctrl-waist-scan}a, the optical control pulse efficiency was simulated as a function of control waist mode for our long crystal length of 12.5 mm, for several angles. From this simulation, it is clear that the efficiency is sensitive to the control mode waist optimization for larger angles. 

\begin{figure}[h]
    \centering
	\includegraphics[scale=0.55]{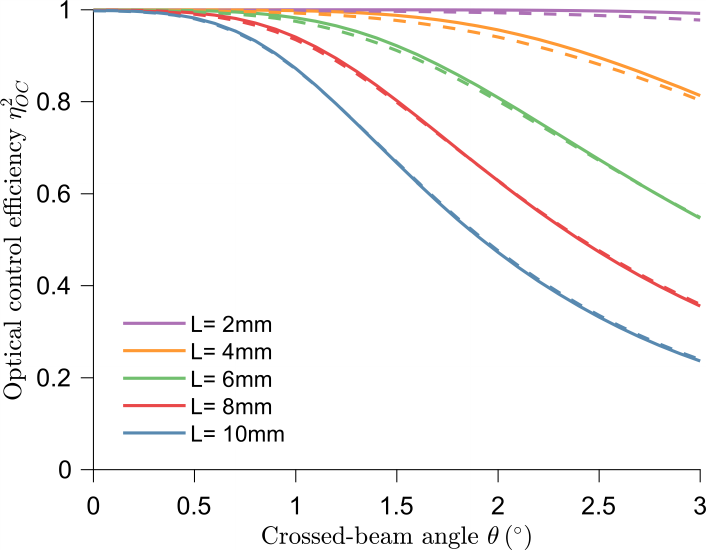}
   \caption{Numerically calculated optical control efficiency $(\eta_{\rm OC}^{\rm th})^2$ as a funciton of crossed-beam angle, for different crystal lengths. For each angle, the optimum control mode waist was found, cf. Fig. \ref{fig:model-ctrl-waist-scan}(a). For the solid lines, the input mode waist was scaled with the crystal length, see Eq. \ref{eq:opt-waist}, for the dashed lines the waist was fixed to 34 $\mu$m. All other simulation parameters were the same as for the simulations shown in Fig. \ref{fig:model-ctrl-waist-scan}.}
	\label{fig:crystal_length}
\end{figure}

To gain further insight into the role of the crystal length, we extracted the maximum efficiency for the optimum control waist, as a function of the crossed-beam angle, for several crystal lengths $L$, as shown in Fig. \ref{fig:crystal_length}. It clearly shows that the better mode overlap in shorter crystals significantly increases the memory efficiency at larger angles for shorter crystals. Note that the choice of the input mode is not critical, as shown in Fig. \ref{fig:model-ctrl-waist-scan}b, the main gain stems from the better mode overlap over a shorter interaction distance. In a cavity-enhanced experiment, there is a competition between crystal absorption and intra-cavity losses, but our simulations here shows that shorter crystals are highly advantageous for reaching high optical-to-spin efficiency for spin-wave memories. 

Finally, the efficiency optimization strategy outlined here is particularly relevant for achieving high signal-to-noise ratio when storing single photon states produced from a spontaneous parametric down conversion (SPDC) source \cite{Seri2017,Rakonjac2021}, a key resource for quantum repeaters \cite{Simon2007}. Building on previous work where weak coherent states at the single photon level were stored in \euyso{} for up to 20 ms, we believe the methods developed here will allow storing entangled SPDC photons on the same timescale in \euyso{}.

\section*{Acknowledgments}

This work has received funding from the Swiss State Secretariat for Education, Research and Innovation (SERI) under contract number UeM019-3. The authors acknowledge initial work and interesting discussions with Dr. Adrian Holz\"{a}pfel.

\appendix
\setcounter{section}{0}
\section{Optical-hyperfine relative transition probabilities}
\label{app:branching-table}

Table A1 shows the relative transition probabilities between the ground and excited state hyperfine levels, calculated using the spin Hamiltonian \cite{ZambriniCruzeiro2018a} for the experimentally estimated magnetic field of $231$ mT at an angle of $6.9^{\circ}$ to the D$_2$ axis.

\begin{figure}[h]
    \includegraphics[scale=0.9]{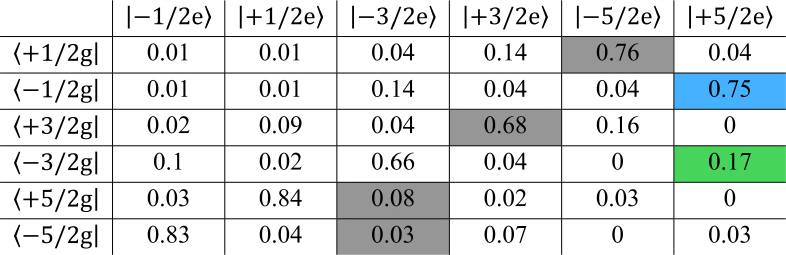}
    \centering
    \caption*{Table A1: Relative transition probabilities for all optical-hyperfine transitions for the specific magnetic field configuration used in this work. The transitions employed in the experiment are indicated using the same color-codes as in Fig. \ref{fig:EuYSOlevels}. The values have been rounded of to the second digit.}
\end{figure}

\section{Input and control transverse mode profiles}
\label{app:mode_profiles}
\begin{figure}[h]
    \centering
	\includegraphics[scale=0.55]{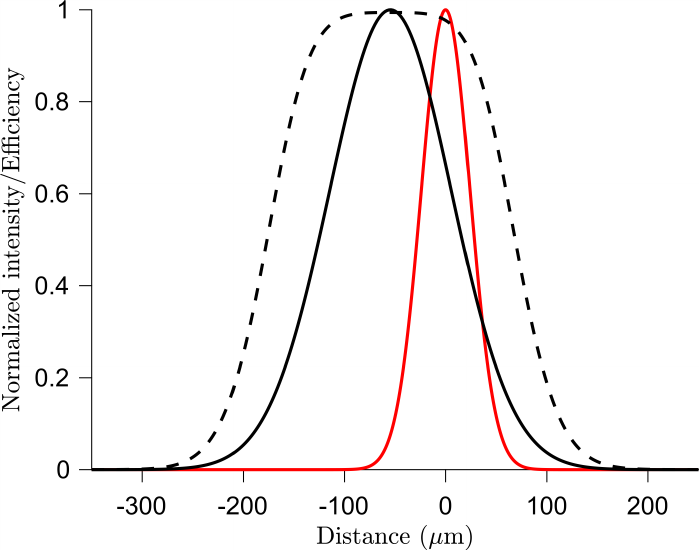}
   \caption{Normalized intensity input (red solid line) and control (black solid line) mode profiles along the $x$ coordinate, for $y=0$ and at the crystal surface $z = L/2$. The dashed black line shows the single pulse efficiency $\eta_{\bar{\delta}}(x,y = 0,z = L/2)$, spectrally averaged over the optical spectrum, see Sec. \ref{sec:num-simul-cbc}. The crossed-beam angle is $\theta = 0.5$ deg.}
	\label{fig:mode_profiles}
\end{figure}

In Fig. \ref{fig:mode_profiles}, we show the input and control intensity mode profiles at the surface of the crystal, for a crossed-beam configuration, along with the single-pulse optical control efficiency profile. The input mode waist is 34 $\mu$m, the control mode waist 120 $\mu$m, and all other parameters correspond to those of Fig. \ref{fig:model-ctrl-waist-scan}a. This illustrates the benefit of an adiabatic pulse, where the pulse efficiency saturates close to unity over a spatial distance significantly wider than the control waist. A longer pulse, or a higher Rabi frequency, further widens the efficiency profile with respect to the intensity mode profile. 

\section{AFC echo efficiency}
\label{app:AFC_echo}

\begin{figure}[h]
    \includegraphics[scale=0.55]{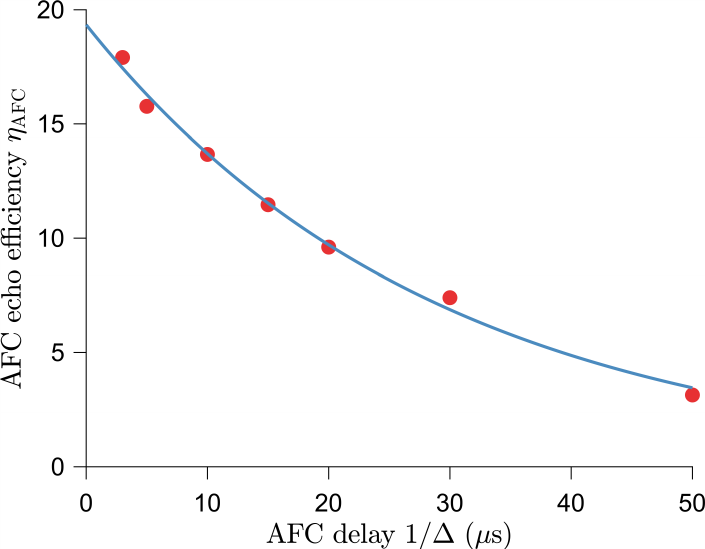}
    \centering
    \caption{AFC echo efficiency decay measurement.}
    \label{fig:AFC_delays}
\end{figure}

The AFC echo efficiency was measured as a function of the $1/\Delta$ delay, while optimizing the AFC creation pulse parameters for each delay, as shown in Fig. \ref{fig:AFC_delays}. The effective AFC coherence time \cite{Jobez2016} is $T_2^{\rm AFC} = (116 \pm 5) \mu$s. Note that this value is significantly lower than the $(300 \pm 30) \mu$s previously achieved in \euyso{} \cite{Ortu2022}, due to higher strain-induced optical decoherence \cite{LouchetChauvet2019} caused by the vibrating pulse tube cooler. The extrapolated zero-delay efficiency is $(19.4 \pm 0.4)\%$, in close agreement with the expected theoretical value of $20.3\%$ for an optimal square-shaped comb with a peak optical depth of $d = 2.4$ \cite{Bonarota2010} (our initial optical depth after the spin polarization step).

\pagebreak

\bibliography{qmcommon}
\bibliographystyle{iopart-num}

\end{document}